\documentclass[%
superscriptaddress,
preprint,
 amsmath,amssymb,
aps,
prl,
]{revtex4-2}

\usepackage{hyperref}
\hypersetup{colorlinks=true,linkcolor=blue,citecolor=blue,urlcolor=blue}
\usepackage{color}
\usepackage{xcolor}
\usepackage{graphicx}
\usepackage{dcolumn}
\usepackage{siunitx}
\usepackage{bm}
\usepackage[version=4]{mhchem}

\usepackage[colorinlistoftodos]{todonotes}

\begin{document}

\title{Multiple Topological States in \texorpdfstring{LaAgAs$_2$}{}, a Failed Square-Net Semimetal} 

\author{Yang Liu$^*$}
\affiliation{Low Temperature Physics Laboratory, College of Physics and Center of Quantum Materials and Devices, Chongqing University, Chongqing 401331, China.
}

\author{Tongrui Li$^*$}
\affiliation{National Synchrotron Radiation Laboratory, University of Science and Technology of China, Hefei 230026, People’s Republic of China.
}

\author{Xixi Yuan$^*$}
\affiliation{College of Physics and Center of Quantum Materials and Devices, Chongqing University, Chongqing 401331, China.}

\author{Nour Maraytta}
\affiliation{Institute for Quantum Materials and Technologies, Karlsruhe Institute of Technology, Kaiserstraße 12, 76131 Karlsruhe, Germany.}

\author{Alexei V. Fedorov}
\affiliation{Advanced Light Source, Lawrence Berkeley National Laboratory, Berkeley, California 94720, USA.}

\author{Asish K. Kundu}
\affiliation{National Synchrotron Light Source II, Brookhaven National Laboratory, Upton, New York 11973, USA.
} 

\author{Turgut Yilmaz}
\affiliation{Department of Physics, Xiamen University Malaysia, Sepang 43900, Malaysia.}

\author{Elio Vescovo}
\affiliation{National Synchrotron Light Source II, Brookhaven National Laboratory, Upton, New York 11973, USA.
} 

\author{Xueliang Wu}
\affiliation{Low Temperature Physics Laboratory, College of Physics and Center of Quantum Materials and Devices, Chongqing University, Chongqing 401331, China.
}

\author{Long Zhang}
\affiliation{Low Temperature Physics Laboratory, College of Physics and Center of Quantum Materials and Devices, Chongqing University, Chongqing 401331, China.
}

\author{Mingquan He}
\affiliation{Low Temperature Physics Laboratory, College of Physics and Center of Quantum Materials and Devices, Chongqing University, Chongqing 401331, China.
}

\author{Yisheng Chai}
\affiliation{Low Temperature Physics Laboratory, College of Physics and Center of Quantum Materials and Devices, Chongqing University, Chongqing 401331, China.
}

\author{Xiaoyuan Zhou}
\affiliation{Low Temperature Physics Laboratory, College of Physics and Center of Quantum Materials and Devices, Chongqing University, Chongqing 401331, China.
}

\author{Michael Merz}
\affiliation{Institute for Quantum Materials and Technologies, Karlsruhe Institute of Technology, Kaiserstraße 12, 76131 Karlsruhe, Germany.}
\affiliation{Karlsruhe Nano Micro Facility, Karlsruhe Institute of Technology, Kaiserstraße 12, 76131 Karlsruhe, Germany.}

\author{Zhe Sun}
\affiliation{National Synchrotron Radiation Laboratory, University of Science and Technology of China, Hefei 230026, People’s Republic of China.
}

\author{Huixia Fu$^\dagger$}
\affiliation{College of Physics and Center of Quantum Materials and Devices, Chongqing University, Chongqing 401331, China.}
\affiliation{Chongqing Key Laboratory for Strongly Coupled Physics, Chongqing University, Chongqing 401331, China.}

\author{Tonica Valla$^\ddagger$}
\affiliation{Donostia International Physics Center, Donostia - San Sebastián, 20018, Spain.
} 
\affiliation{Institut za fiziku, Bijeni\v{c}ka 46, HR-10000 Zagreb, Croatia
}

\author{Aifeng Wang$^\S$}
\affiliation{Low Temperature Physics Laboratory, College of Physics and Center of Quantum Materials and Devices, Chongqing University, Chongqing 401331, China.
}

\collaboration{*These authors contributed equally to this work.
$^\dagger$Corresponding author: hxfu@cqu.edu.cn;
$^\ddagger$Corresponding author: tonica.valla@dipc.org; 
$^\S$Corresponding author: afwang@cqu.edu.cn.}

\date{\today}

\begin{abstract}
The rational design of new materials emerges as an important direction to explore new topological materials, which is based on the understanding of the correlation between crystal and electronic structures. In this paper, we perform a comprehensive study on the crystal and electronic structures in \ce{LaAgAs2} through a combination of single-crystal x-ray diffraction (XRD), quantum oscillation, and angle-resolved photoemission spectroscopy (ARPES) experimental measurements, and density functional theory (DFT) calculations. Single-crystal XRD measurements reveal that \ce{LaAgAs2} crystallizes into a \ce{HfCuSi2}-derived structure with the square net distorted into cis-trans chains. Quantum oscillation measurements reveal two frequencies with small effective masses and quasi-two-dimensional (2D) characters. ARPES measurements reveal an electronic structure strikingly different from the square-net-based semimetals, such as \ce{LaAgSb2}. The Fermi surface is quasi-two-dimensional (2D), with Dirac-like hole pockets at the zone center and a quasi-1D elliptical electron pocket at the zone boundary. Based on the DFT calculations, the measured electronic structure can be well understood regarding the cis-trans distortion, which transforms the two-dimensional square net-derived Dirac bands into quasi-1D trivial bands. Intriguingly, multiple topological states can be identified around the zone center, including a nontrivial $Z_{2}$ topological surface state and a bulk Dirac state. Our study clarifies the impact of cis-trans distortion and identifies \ce{LaAgAs2} as a topological material with multiple topological states near the Fermi level, providing a guideline for intentionally designing new topological materials.
\end{abstract}

\maketitle

\section{Introduction}

Topological materials, including topological insulators and topological semimetals, are unconventional phases of matter characterized by topologically nontrivial band structures \cite{kumar_topological_2021,ando_topological_2013,armitage_weyl_2018}. Owing to their unconventional electronic and transport properties, great effort has been devoted to the discovery of new topological materials \cite{wieder_topological_2021,vergniory_all_2022,cano_band_2021,xu_high-throughput_2020,schleder_dft_2019}. One of the most successful strategies is to design the new layered materials using the LEGO-like building block approach, specifically by alternatively stacking structural motifs that host desired band structure and the buffer layers along the out-of-plane direction \cite{klemenz_systematic_2020,klemenz_topological_2019,jovanovic_simple_2022,wang_quantum_2023}. This strategy is based on the assumption that the electronic structure of the building block can be regarded as an entirety, which could keep its main feature when embedded in real materials \cite{hoffmann_how_1987,tremel_square_1987,hoffmann_making_1985}. However, in real materials, both the crystal and band structures of the building blocks are tunable by local chemical environments, for example, structural distortions and band dispersion. To improve the efficiency and accuracy of designing new topological materials, it is important to clarify the impact of structural distortions and explore new building blocks.

Due to its simplicity and prevalence in real materials, the planar square lattice has attracted extensive theoretical and experimental interest \cite{klemenz_topological_2019}. Theoretically, it has been extensively studied in the context of the Ising model and 2D Dirac semimetal \cite{onsager_crystal_1944,young_dirac_2015}. Experimentally, the square lattice is found to be linked to various intriguing phenomena, such as anisotropic Dirac fermions \cite{lee_anisotropic_2013,feng_strong_2015}, charge density waves (CDWs) \cite{malliakas_square_2005,wu_coexistence_2023}, and superconductivity \cite{akiba_observation_2022}, making it an excellent platform for scrutinizing the intertwined orders. Notably, cuprate-, iron-, and nickel-based high-temperature superconductors with complex phase diagrams are associated with the square lattice \cite{park_structures_1995,fernandes_iron_2022,sun_signatures_2023,Valla2020a}. In real materials, the square lattice might undergo structural distortions, e.g., distorted into zigzag chains or cis-trans trains, which will alter its electronic structure and related physical properties \cite{hulliger_gdps_1977}. Therefore, clarifying the distortion effects on the square lattice is crucial to understanding square-net-based materials.

In topological materials, many crystal structures can host the prototype square-net-derived band structure, which can be classified as the square-net-based topological materials \cite{lee_topological_2021,klemenz_topological_2019}. Primary examples include the \ce{HfCuSi2}-type structure (e.g., \ce{SrMnBi2} and \ce{LaAgSb2} \cite{park_anisotropic_2011,wang_multiband_2012}) and PbFCl-type structure (e.g., \ce{ZrSiS} and \ce{GdSbTe} \cite{schoop_dirac_2016,venkatesan_direct_2025}), as well as \ce{PtPb4} \cite{wu_nonsymmorphic_2022}. Owing to their structural diversity and the robust square-net-derived bands, \ce{HfCuSi2}-type compounds with the general chemical formula $AMPn_2$ ($A$ = alkaline earth or rare earth; transition metal; $Pn$ = Sb or Bi) are a central platform for studying the square-net-based topological semimetals \cite{klemenz_topological_2019}. The crystal structure is typically described as the alternating stacking of the $A$ layer, anti-PbO-type [$MPn$] layer, and square-net-type $Pn$ planar layer along the stacking direction. Most of the notable phenomena observed in $AMPn_2$ materials arise from the $Pn$ square lattice. $A$ and [$MPn$] layers are merely treated as buffer layers that influence physical properties through interaction with the $Pn$ square lattice \cite{lee_anisotropic_2013}. 

Previous studies mainly focus on tuning the band structure of $Pn$ square net via manipulating the $A$ and [$MPn$] layers \cite{yang_single-crystal_2023,xia_coupling_2023}. However, little attention has been paid to the buffer layers and the distortion of the square net. However, it has been reported that the [FeAs]/[FeTe] layer with the same crystal structure as the [$MPn$] layer could host multiple topological states \cite{zhang_multiple_2019}. Hoffmann \emph{el. al.} theoretically pointed out that $Pn$ atoms in [$MPn$] layer are also arranged in the square net fashion, potentially showing the characteristic square-net bands \cite{hoffmann_making_1985,tremel_square_1987}. Moreover, it has been theoretically demonstrated that the Peierls-like distortions could significantly modify the square net derived band structure \cite{tremel_square_1987}. The possible distortion of the square net mainly includes the zigzag chains and cis-trans chains. Zigzag chains have been often observed among \ce{$A$MnSb2} ($A =$ Sr, Ba, Eu) compounds, showing a similar band structure with the undistorted square nets \cite{rong_electronic_2021}. However, experimental validation is still lacking for the cis-trans distortion due to the lack of suitable materials. Therefore, it is of great interest to systematically study the topological properties of buffer layers and the impact of cis-trans distortion. 

Recently, ternary rare-earth coinage metal antimonide \ce{La$T$Sb2} ($T =$ Cu, Ag, and Au) has gotten renewed interest due to the coexistence of charge density wave (CDW), Dirac fermion, and superconductivity \cite{wu_coexistence_2023,akiba_observation_2022}. \ce{LaAgAs2} shares a similar crystal structure with \ce{La$T$Sb2} but with the planar layer distorted from the Sb square net into As cis-trans chains \cite{rutzinger_lattice_2010}, which is expected to show similar physical properties. However, previous studies mainly focus on the crystal structure \cite{demchyna_new_2001,eschen_preparation_2003,rutzinger_lattice_2010}, with the electronic structure and physical properties remaining elusive, possibly due to the lack of high-quality single crystals. 

In this study, we successfully synthesized high-quality \ce{LaAgAs2} single crystals using the self-flux method. Through comprehensive crystal and electronic structure measurements, our experimental findings reveal that rather than being a square-net-based topological semimetal with symmetry-induced band inversion between As1 $p_x/p_y$ bands, \ce{LaAgAs2} is identified as a topological material with multiple topological states.

\section{RESULTS}

\subsection{Sample characterization}

\begin{figure*}[htp!]
\includegraphics[width = \linewidth]{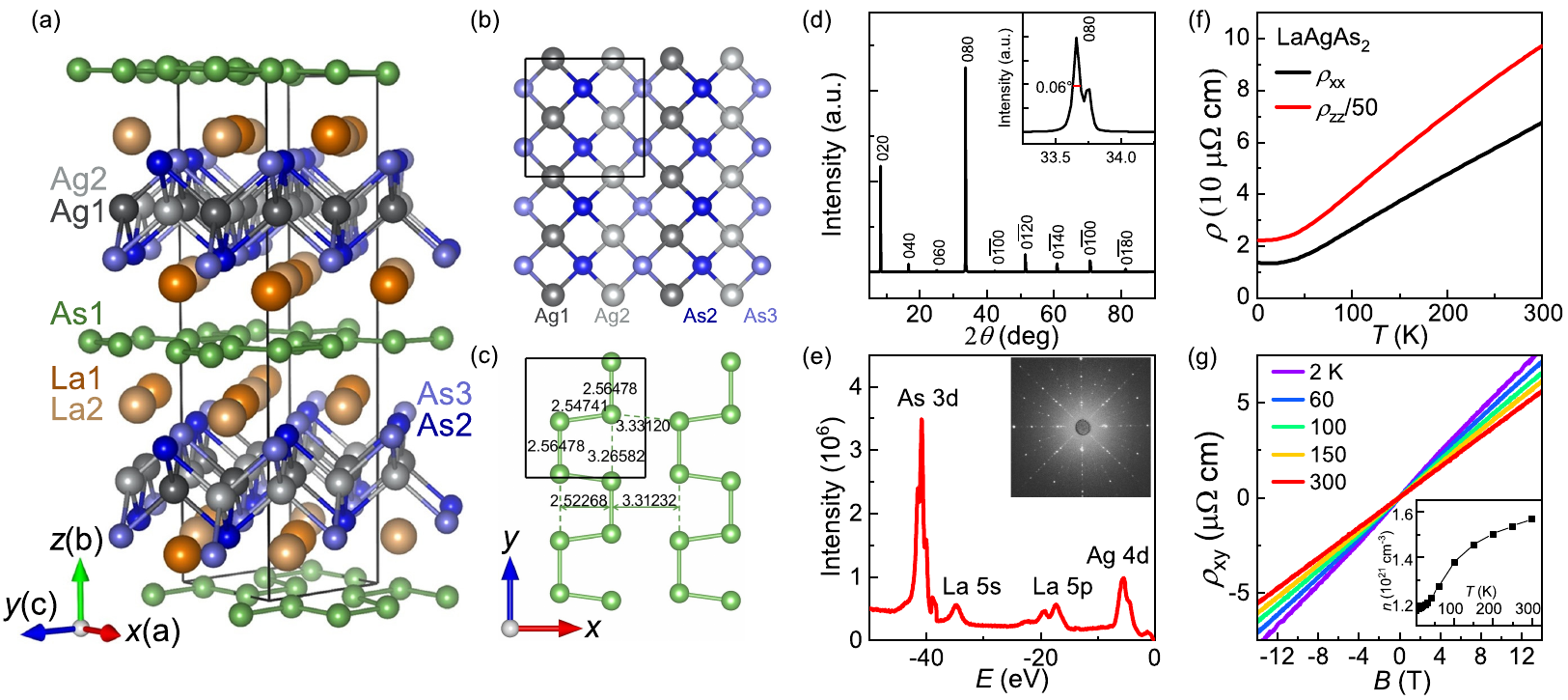}
\caption{Crystal structure and basic physical properties of \ce{LaAgAs2}. (a) Crystal structure of \ce{LaAgAs2}. To avoid confusion, the crystallographic direction for $a$, $b$, and $c$ were denoted as $x$, $z$, and $y$, respectively. Top view of [AgAs] layer (b) and As planar layer (c), where panels (b) and (c) use the same coordinate axis as shown in panel (c). Black lines in panels (a-c) indicate the unit cell of \ce{LaAgAs2}. (d) XRD pattern for (0 $k$ 0) surface of a flat \ce{LaAgAs2} crystal.  The insert shows the enlarged view of the (0 8 0) reflection. (e) Core-level electronic structure of \ce{LaAgAs2}, measured using a photon energy of 150 eV. The inset shows an x-ray Laue pattern of the (H, 0, L) reciprocal plane. (f) Temperature dependence of in-plane resistivity $\rho_{xx}$ ($j \parallel ac$) and out-of-plane resistivity $\rho_{zz}$ ($j \parallel b$) measured with $B$ = 0 T for LaAgAs$_2$, where $j$ is the electric current applied for the resistivity measurements. (g) Typical Hall resistivity $\rho_{xy}(B)$ curves measured at different temperatures with $B \parallel b$. The insert shows the temperature dependence of carrier concentration obtained from the single-band fitting of the high-field part of $\rho_{xy} (B)$ curves. 
}
\label{fig1}
\end{figure*}

The single-crystal XRD measurements revealed that \ce{LaAgAs2} crystallizes in an orthorhombic crystal structure (space group: $Pbcm$) with lattice parameters $a$ = 5.8350(1) {\AA},\@ $b$ = 21.2926(4) {\AA},\@ and $c$ = 5.8306(1) {\AA},\@ consistent with those reported by Rutzinger \emph{et.\@ al.\@} \cite{rutzinger_lattice_2010}.\@ The obtained crystal structure is shown in Fig. \ref{fig1}(a-c) and the refined parameters are listed in Table \ref{tab_cryst}.\@ The crystal structure of LaAgAs$_2$ [Fig.\@ \ref{fig1}(a)] consists of alternatively stacking the anti-PbO type [AgAs] layer [Fig.\@ \ref{fig1}(b)] and As1 cis-trans chains [Fig.\@ \ref{fig1}(c)] along the $b$ axis, interspersed with La layers. The orthorhombic distortion primarily affects the planar As1 layer, which deforms from a square-net configuration into cis-trans chains. In contrast, the La and [AgAs] layers retain a pseudo-tetragonal structure with an $a/c$ ratio of 1.00075. 

The out-of-plane x-ray diffraction pattern shown in Fig.\@ \ref{fig1}(d) confirms the pure phase of the crystals with a preferred [010] orientation. The narrow peak width [inset of Fig.\@ \ref{fig1}(d)], along with the bright and sharp Laue diffraction spots [inset of Fig.\@ \ref{fig1}(e)], demonstrates the excellent crystallinity nature of the \ce{LaAgAs2} crystals. Laue diffraction exhibits a four-fold symmetry, attributable to the combination of pseudo-tetragonal symmetry and the twinning effect, which will be discussed in detail in the following sections. Figure \ref{fig1}(e) shows the shallow core levels, indicating that only the constituent elements (La, Ag, and As) were present, with no additional peaks. These results suggest that our \ce{LaAgAs2} crystals are phase pure and of high quality. 

As shown in Fig. \ref{fig1}(f), both the in-plane resistivity ($\rho_{xx}$) and out-of-plane resistivity ($\rho_{zz}$) of \ce{LaAgSb2} exhibit a typical metallic behavior across the entire temperature range, in contrast with the semiconducting behavior reported by Rutzinger \emph{et al.} \cite{rutzinger_lattice_2010}. This difference might be attributable to the influence of the grain boundaries in polycrystal pellets. The residual resistivity ratios (RRR) for both $\rho_{xx}$ and $\rho_{zz}$ is about 5. The room-temperature resistivity values are determined to be $\rho_{xx}$(300 K) = \SI{68}{\micro\ohm\centi\metre} and $\rho_{zz}$(300 K) = \SI{4.86}{\mohm\centi\metre}, yielding a large resistive anisotropy ratio of $\rho_{zz}/\rho_{xx} \sim$ 70. This value is approximately an order of magnitude larger than that in \ce{LaAuSb2} \cite{wu_coexistence_2023}, reflecting the strong 2D character for the underlying electronic structure in \ce{LaAgAs2}. 

As shown in Fig. \ref{fig1}(g), the Hall resistivity $\rho_{xy}(B)$ exhibits an approximately linear behavior with a positive slope up to 14 T. A subtle change in slope around $B$ = 4 T can be recognized for temperatures below 100 K, indicative of multiband behaviors dominated by hole-type carriers. Due to the nearly linear characteristics of the $\rho_{xy}(B)$ curves, accurate fitting using a two-band model is challenging. Consequently, the carrier concentration is estimated by single-band linear fitting of the high-field region of the $\rho_{xy}(B)$ curves \cite{zhang_strong_2021}. The resulting carrier concentration, presented in the inset of Fig. \ref{fig1}(g), gradually decreases with decreasing temperature. The carrier concentration at 300 K is estimated to be $n =$ \SI{1.57e21}{cm^{-3}}, equivalent to 0.15 holes per formula unit (f.u.), which is comparable to that in \ce{LaAgSb2} \cite{wang_multiband_2012}.  

\subsection{Quantum oscillations}

\begin{figure*}
\includegraphics[width = \linewidth]{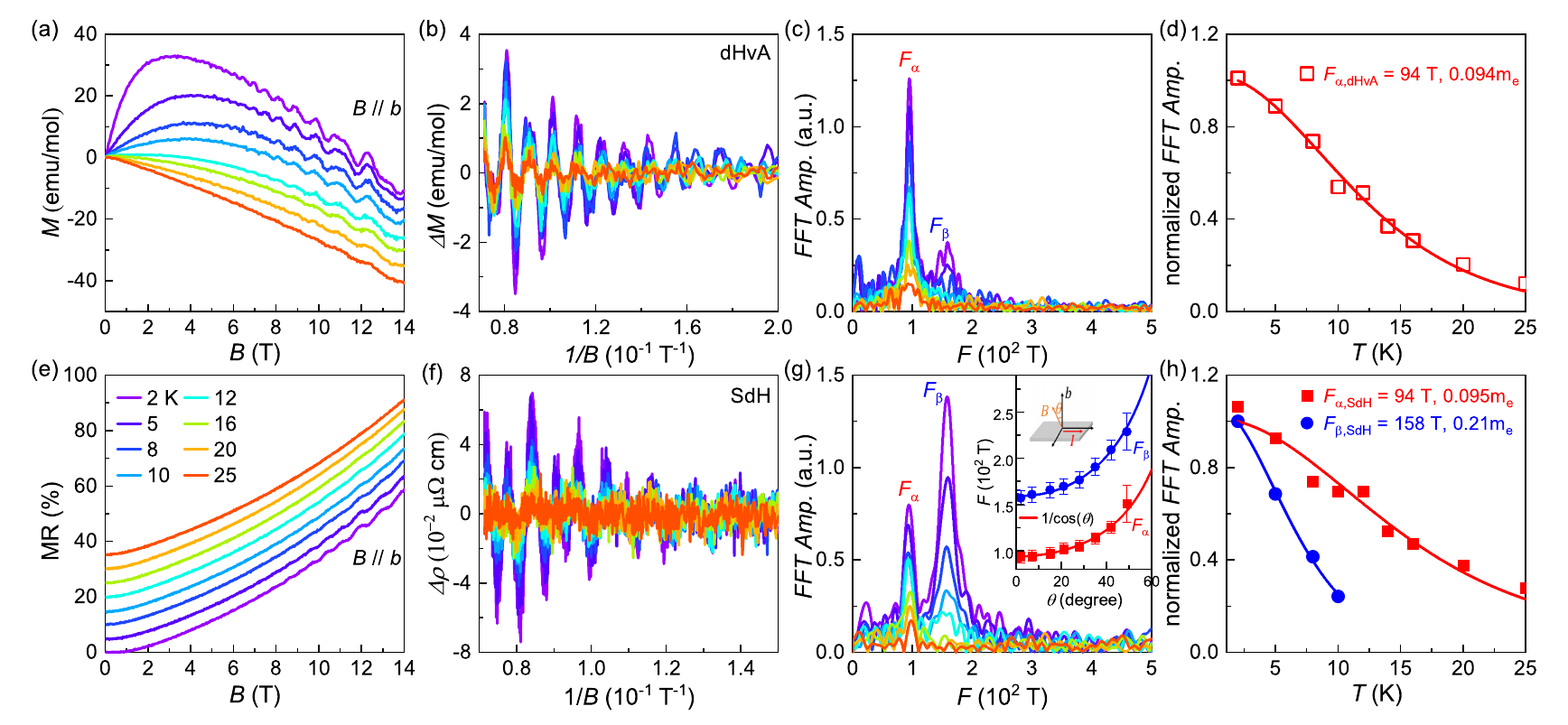}
\caption{Quantum oscillations in \ce{LaAgAs2}. (a)(e) Magnetic field dependence of magnetization [$M(B)$] and magnetoresistance [MR($B$)], respectively. Each subsequent MR curve is shifted upward by 4\% for clarity. (b)(f) Inverse field dependence of oscillatory components of magnetization (${\Delta}M$, dHvA oscillations) and magnetoresistance (${\Delta}{\rho}$, SdH oscillations), respectively. ${\Delta}M$/${\Delta}{\rho}$ are obtained by subtracting the polynomial background from the data in (a)/(e), respectively. (c)(g) FFT spectra of dHvA/SdH oscillations at various temperatures. The inset of (g) shows the angular dependence of FFT peaks and the geometry of the measurements, where $\theta$ is defined as the angle between the magnetic field $B$ and the crystallographic $b$-axis. Further details for the angular dependence of SdH oscillation can be found in Supplementary Fig. S2 \cite{SM}. (d)(h) Temperature dependence of the FFT amplitude for $F_\alpha$ and $F_\beta$, which are the FFT peaks inferred from (c) and (g), respectively. Solid lines represent the fits with the Lifshitz-Kosevich formula.
}
\label{fig2}
\end{figure*}

As one of the most powerful experimental techniques that can probe the band topology, quantum oscillation has been widely adopted to study topological materials \cite{hu_transport_2019,zhao_berry_2022}. In high magnetic fields, the quantization of energy levels into discrete Landau levels leads to periodic oscillations in physical quantities as a function of $1/B$ \cite{shoenberg_1984}. These quantities include magnetization and resistivity, corresponding to de Haas-van Alphen (dHvA) oscillation and Shubnikov-de Haas (SdH) oscillation, respectively. The quantum oscillation signal of \ce{LaAgAs2} is probed by both dHvA and SdH oscillation measurements, with the results presented in Fig. \ref{fig2}. 

Shown in Fig. \ref{fig2}(a) is the isothermal out-of-plane ($B \parallel b$) magnetization measured at various temperatures for a \ce{LaAgAs2} single crystal. Clear oscillation signals are observed on the paramagnetic/diamagnetic background for $B >$ 5 T. The oscillatory components ${\Delta}M$, extracted by subtracting a smooth polynomial background, are plotted as a function of $1/B$ in Fig. \ref{fig2}(b), where a subtle beat pattern can be identified. As shown in Fig. \ref{fig2}(c), fast Fourier transformation (FFT) analyses of ${\Delta}M$ reveal two oscillation frequencies, i.e., $F_\alpha$ = 94 T and $F_\beta$ = 158 T. Figure \ref{fig2}(e) shows the magnetic field dependence of MR, defined as MR = $[\rho(B) - \rho(0)]/\rho(0) \times 100\%$. The MR exhibits a power-law dependence on $B$ (MR $\propto B^n$, $n =$ 1.45 $\sim$ 1.6), reaching a maximum value of 58\% at 2 K and 14 T. SdH oscillation signal superimposed on the MR curves can be seen for $B >$ 7 T, where the oscillating magnitude is weaker than that in magnetization. After the subtraction of a smooth polynomial background, the FFT analysis of $\Delta{\rho}$ in Fig. \ref{fig2}(f) gives rise to the FFT spectrum displayed in Fig. \ref{fig2}(g), revealing two oscillation peaks with nearly identical frequencies to those in the dHvA spectrum [Fig. \ref{fig2}(c)]. However, from the FFT spectra shown in Figs. \ref{fig2}(c) and \ref{fig2}(g), it can be seen that the relative amplitude between $F_\alpha$ and $F_\beta$ is reversed between dHvA and SdH oscillations, which is a common phenomenon and is usually interpreted in terms of different scattering mechanisms \cite{muller_determination_2020,zhang_comprehensive_2024}. Further details for the comparison of the quantum oscillation in $M$, $\rho_{xx}$, and $\rho_{xy}$ can be found in Supplementary Fig. S1 \cite{SM}.

The geometry of the Fermi surface can be inferred from Quantum oscillation measurements via the Onsager relation \cite{shoenberg_1984}: $F=\left(\frac{\Phi_0}{2\pi^2}\right)A_{\mathrm{F}}$, where $\Phi_0$ = \SI{2.07e-15}{Tm^2} is the magnetic flux quantum, and $A_\mathrm{F}$ is the cross-sectional area of the Fermi surface normal to the magnetic field. The cross-sectional area of the Fermi surfaces associated with $F_\alpha$ and $F_\beta$ are calculated to be $A_\mathrm{F,\alpha}$ = \SI{0.90}{nm^{-2}} and $A_\mathrm{F,\beta}$ = \SI{1.51}{nm^{-2}}, respectively, corresponding to 0.8\% and 1.3\% of the total area of the $ac$ plane in the Brillouin zone. From the inset in Fig. \ref{fig2}(g), it can be seen that both $F_\alpha$ and $F_\beta$ show a $\sim 1/\mathrm{cos}(\theta)$ dependence on the orientation of the magnetic field, which is a characteristic of the 2D Fermi surface. These results indicate that $F_\alpha$ and $F_\beta$ originate from quasi-2D Fermi pockets with small sizes.

The effective mass $m = m^*m_e$ is related to the band dispersion around the Fermi level ($E_\mathrm{F}$) via $m^*=\hbar^2 /\left[\partial^2 E\left(k\right) / {\partial}k^2\right]$, which can be extracted by fitting the temperature dependence of the oscillation amplitudes to the thermal damping factor $R_\mathrm{T}$. According to the Lifshitz-Kosevich model, $R_\mathrm{T}$ is defined as \cite{shoenberg_1984}:  
\begin{equation}
R_{\mathrm{T}}=\frac{\alpha T m^{*} / \overline{B}}{\sinh \left(\alpha T m^{*} / \overline{B}\right)} \\
\end{equation} 
where $\alpha = 2\pi^2 k_{\mathrm{B}} m_e / e \hbar=14.69 \mathrm{~T} / \mathrm{K}$ is a constant, $\overline{B} = 1 /\left[\left(1 / B_{\max }+1 / B_{\min }\right) / 2\right]$ is the average inverse field used in the FFT analysis. As shown in Fig. \ref{fig2}(d), the FFT amplitudes vs temperature curves can be well-fitted by $R_\mathrm{T}$, yielding effective masses of $m_\alpha = 0.094m_e$ and $m_\beta = 0.21m_e$. These effective masses are comparable to those in sisiter compounds such as \ce{$RE$AgSb2} [(0.07 $\sim$ 0.5)$m_e$] \cite{myers_haasvan_1999}, and \ce{$A$MnSb2} [(0.05 $\sim$ 0.1)$m_e$] \cite{liu_magnetic_2017,xia_coupling_2023}, as well as \ce{La3ScBi5} ($\sim$0.2 $m_e$) \cite{xu_quasi-linear_2025}, where $RE$ and $A$ represents the rare-earth and alkaline-earth elements, respectively. The small value of effective masses in \ce{LaAgAs2} indicates strong band dispersions for both the $F_\alpha$ and $F_\beta$ pockets, possibly linear Dirac bands as those reported in sister $AMPn_2$ compounds \cite{wu_coexistence_2023}. 

The general behavior of quantum oscillations for \ce{LaAgAs2} is similar to the well-established square-net-based topological semimetals such as \ce{LaAgSb2}, \ce{LaAuSb2}, \ce{SrMnSb2}, and \ce{EuMnSb2} \cite{myers_haasvan_1999,wu_coexistence_2023,liu_magnetic_2017,zhang_strong_2021}, indicating a similar band structure and topological properties at first glance. However, the detailed band structure and the origin of $F_\alpha$ and $F_\beta$ remain elusive. Therefore, the electronic structure of \ce{LaAgAs2} is further studied by ARPES, a technique that maps the band structure directly.

\subsection{ARPES}

\begin{figure}
\centering
\includegraphics[width=8cm]{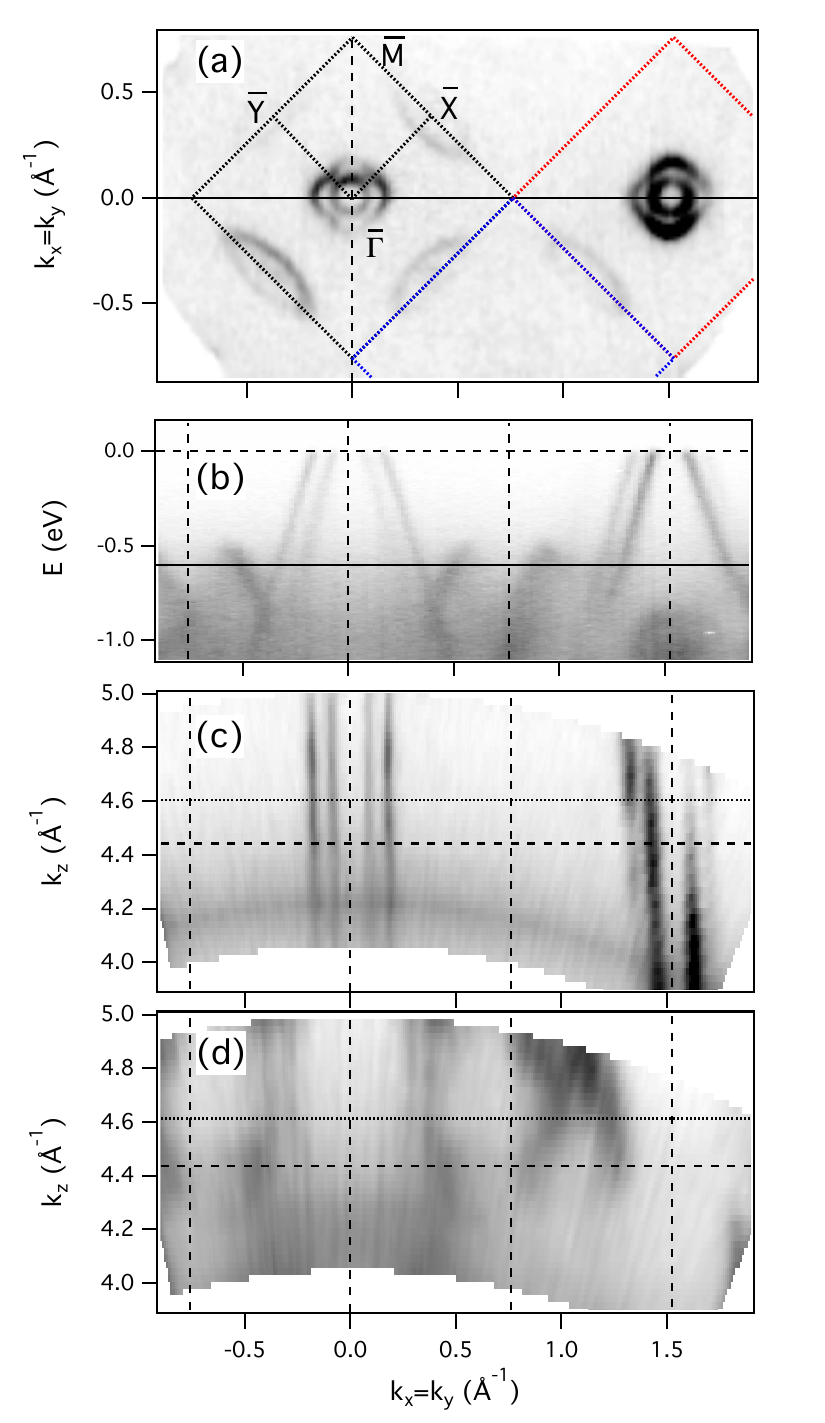}
\caption{Electronic structure of LaAgAs$_2$ from ARPES. (a) The Fermi surface is taken at $h\nu=70$ eV. The black, blue, and red dotted squares represent the $1^{st}$, $2^{nd}$, and $3^{rd}$ surface BZs, respectively, with the symmetry points indicated. (b) The $E(k)$ dispersion along the $\bar{\Gamma}-\bar{\mathrm{M}}$ line of the surface BZ (solid horizontal line in (a)). (c) The $k_z$ dependence of the states at the Fermi level ($E=0$) along the same momentum line. (d) The same as in (b), but at $E=-0.6$ eV marked by the black solid line in (b). The maps in (c, d) are obtained by using the photon energies in the range from 55 to 90 eV and the free-electron approximation for the final electron state, $k_z=1/\hbar\sqrt{2m_e(E_kcos^2(\theta)+V_0)}$, where $E_k$ is the kinetic energy of a photoelectron and $V_0\sim10$ eV is the inner potential. The dashed and dotted horizontal lines mark the $\Gamma$ and Z planes in the 16$^{th}$ 3D BZ, respectively. All the spectra were taken at $T$ = 15 K using the horizontally linearly polarized light.
}
\label{fig3}
\end{figure}

Figure \ref{fig3} shows the in-plane Fermi surface (a) and the band dispersions along the BZ diagonal (b), taken with $h\nu=70$ eV photons. The photon-energy dependence that gives access to the out-of-plane electronic structure is shown in panels (c) and (d). The set of spectra along the zone diagonal, $\bar{\Gamma}-\bar{\mathrm{M}}-\bar{\Gamma}$, measured at photon energies ranging from 55 to 90 eV is converted to $k_z$ using the free electron approximation for the final photoemission state, with $V_0=10$ eV for the "inner" potential. Panel (c) shows the intensity at the Fermi level (Fermi surface), while panel (d) shows intensity at $E=-0.6$ eV. Almost total absence of dispersion in (c) indicates a highly 2D character of the states forming the Fermi surface, whereas weakly dispersing contours in (d) indicate that deeper states have a certain 3D character. 

\begin{figure*}
\includegraphics[width=14cm]{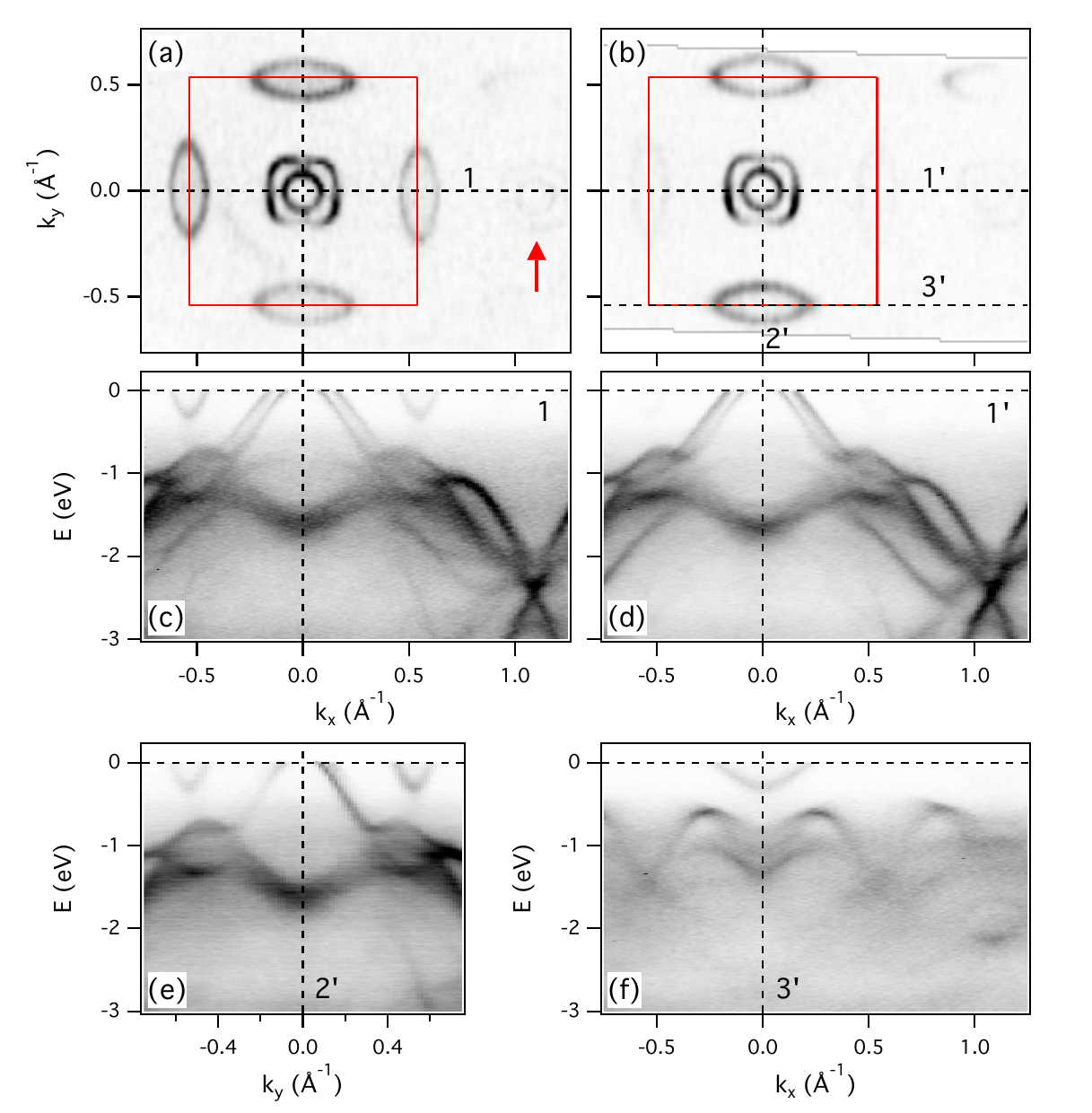}
\caption{In-plane electronic structure of LaAgAs$_2$. (a) The Fermi surface from the spot with 2 orthogonal domains. (b) The Fermi surface from the spot with a single domain dominating. (c) Band structure along the $k_y=0$ (marked as 1) line from (a). (d) The same as (b) (marked as 1'). (e) Band structure along the $k_x=0$ (marked as 2') line from (b). (f) Band structure along the $k_y=\pi/c$ (marked as 3') line from (b). The red arrow points to the suppressed intensity of the $2^{nd}$ zone states. The red square in (a,b) represents the $1^{st}$ BZ. All the spectra were taken at $T=15$ K using the horizontally linearly polarized light at 100 eV.
}
\label{fig4}
\end{figure*}

In Fig. \ref{fig4} we show the more detailed in-plane electronic structure, recorded at $h\nu=100$ eV. Panel (a) shows the Fermi surface from the spot on the sample that has two orthogonal orthorhombic domains contributing nearly equally to the ARPES intensity. In panel (b), recorded only $\sim20$ $\mu$m from the spot probed in (a), one domain clearly dominates. Panel (c) shows the spectrum from the $k_y=0$ ($\bar{\Gamma}-\bar{\mathrm{Y}}$) line of the surface BZ of the twinned spot (a), while panels (d-f) show the spectra along the three different momentum lines, as indicated, recorded from the single-domain spot (b). 

There are several important findings that can be immediately deduced from Fig. \ref{fig4}. First, consistent with our XRD results, it is obvious that the crystals of \ce{LaAgAs2} are twinned. The domains are large enough that they can be completely resolved in ARPES at both beamlines, indicating that they are of the order of tens of microns across. Obviously, the twinning will affect the macroscopic transport properties measured on the twinned crystals and remove any anisotropy related to the quasi-1D character of the electron pockets visible in Fig. \ref{fig4}(b). Indeed, as shown in Supplementary Fig. S3 \cite{SM}, with the magnetic field being rotated in the $ac$ plane, the angular dependence of MR exhibits a four-fold symmetry instead of two-fold symmetry as would be expected for the quasi-1D electron pockets. 

Therefore, the states at the Fermi level, that determine transport properties, are either quasi-2D (holes) or quasi-1D (electrons), with almost no out-of-plane dispersion, in good agreement with the large $\rho_{zz}/{\rho_{xx}}$ ratio and the quantum oscillation measurements [inset of Fig. \ref{fig2}(g)]. The $\Gamma$-centered hole states disperse almost linearly within the planes, with relatively high Fermi velocities, indicating a very light character in good agreement with magnetotransport. This, and the absence of significant broadening with energy or temperature, should be reflected in very high hole mobilities. 

Our high-resolution measurements show that the larger contour actually consists of two very close pockets, as can be seen in Fig. \ref{fig5}(a-c) (additional details of both the hole and electron pockets can be seen in Supplementary Fig. S8 \cite{SM}). In the inner, circle-like shaped contour, the two states predicted by calculations cannot be resolved. The area of the Fermi surface, which gives the concentration of carriers through the Luttinger count, can be directly measured from the positions of momentum distribution curves (MDC) peaks, $k_F$ \cite{Valla1999a,Valla2020a}. The inner and outer hole contours areas recorded at $h\nu=100$ eV are $A_i=2.3$ nm$^{-2}$ and $A_o=8.6$ nm$^{-2}$, respectively. If recorded at $h\nu=70$ eV [Fig. \ref{fig5}(d)], both hole contours enclose slightly smaller areas, $A_i=1.6$ nm$^{-2}$ and $A_o=7.3$ nm$^{-2}$, respectively, indicating a small, but finite $k_z$ warping, in excellent agreement with the DFT calculations. The smaller one, $A_i$, is very close to the larger orbit $F_{\beta}$ in the dHvA oscillations. The estimated Fermi velocities of the inner and outer hole bands along the $\bar{\Gamma}-\bar{\mathrm{X}}$ and $\bar{\Gamma}-\bar{\mathrm{Y}}$ lines, determined from MDC derived dispersions, are almost the same, $v_F\approx2.5$ eV{\AA} ~(3.8$\times10^5$ ms$^{-1}$). The only exception is one of the states forming the outer doublet that disperses somewhat faster: $v_F\approx3.6$ eV{\AA} ~(5.5$\times10^5$ ms$^{-1}$). 
In the $\bar{\Gamma}-\bar{\mathrm{M}}$ direction all holes have nearly the same Fermi velocity, $v_F\approx 3.3$ eV{\AA} ~(5$\times10^5$ ms$^{-1}$). Taking all the hole pockets into account, and their quasi-2D character, we can estimate the concentration to be $\approx0.34$ holes/f.u. 

The electron pockets around $\bar{\mathrm{X}}$ points are the only indication of a structurally quasi-1D character of the crystal, originating from the cis-trans chains of As1 atoms inside the As1 planar planes. These pockets show more conventional parabolic dispersion. Their Fermi surface encloses an area $A_{el}=5.3$ nm$^{-2}$, somewhat smaller than in DFT calculations. Further disagreement with the DFT is visible at $h\nu=70$ eV: a very small electron-like band is just touching the Fermi level, enclosing a tiny electron pocket inside the big one [Fig. \ref{fig5}(d,e)]. This state is not present in the DFT calculations. Its area cannot be precisely determined in ARPES, but it could range from 0.5 to 1 nm$^{-2}$. 

The Fermi velocity of the main electron pocket is $v_F\approx4$ eV\AA ~(6.1$\times10^5$ ms$^{-1}$) and $v_F\approx1.9$ eV\AA ~(2.9$\times10^5$ ms$^{-1}$) along the $\bar{\Gamma}-\bar{\mathrm{X}}$ and perpendicular to it, respectively. Corresponding effective masses are 0.07 $m_0$ and 0.53 $m_0$, respectively. The electron concentration from this pocket is estimated to be 0.18 electrons per site, by assuming that it too is a doublet and that it is perfectly 2D. The resulting excess hole concentration is then ~0.16 per site, in perfect agreement with the Hall coefficient [Fig. \ref{fig1}(g)]. 

Further, there is a peculiar intensity modulation of photoelectron intensity near the Fermi level measured at different BZs: strong intensity from the $1^{st}$ BZ, almost vanishing in the $2^{nd}$ and again strong in the $3^{rd}$ (Fig.\ref{fig3}(a)). The diminishing intensity in the $2^{nd}$ BZ points to some sort of destructive interference, probably related to the cis-trans structure motif of the As1 chains. Also, the states further away from the Fermi level, $E\leq-0.6$ eV, shown in Fig. \ref{fig4}(c-f), do not show the same periodicity as the As1-derived states forming the Fermi surface. These deeper states are mainly the [AgAs]-derived states that keep the original tetragonal symmetry and due to the weak coupling with the As1 layers, do not repeat in the orthorhombic BZ.

Finally, we note that according to the crystal structure and calculated exfoliation energies for possible cleavage surfaces [see Supplementary Fig. S4 \cite{SM}], there should be at least two different terminations when the crystal is cleaved. Due to the very shallow probing depth, ARPES should be sensitive to that. Our calculations show that the La--As1 interface has the lowest exfoliation energy of the three possible interfaces within the crystal. This suggests that the single crystal preferentially exposes the La and the As1 atomic layers upon cleaving. In reality, the actual surface will likely be a mixture of these two terminations \cite{Gibson2013a}. The fact that ARPES could not distinct between different terminations would indicate that 1) the lateral size of each termination domain is smaller than our spot size, or 2) that the spectroscopic difference between the terminations is not significant, or 3) some reconstruction occurs, leading to a surface that is crystallographically and chemically distinct from bulk, but uniform across the surface.

\subsection{Electronic structure calculations}

DFT calculations were performed to understand the electronic structure experimentally detected by quantum oscillation and ARPES measurements. From the density of states (DOS) shown in Fig. \ref{fig6}(a), it is evident that the Fermi level of \ce{LaAgAs2} is dominated by the states of La and As(2,3), differing from the square-net structured $Pn$ layer in $AMPn_2$ materials \cite{wang_multiband_2012,wu_coexistence_2023,xia_coupling_2023,islam_controlling_2020}. Further in-depth DOS analysis can be seen in Supplementary Figs. S5 and S6 \cite{SM}. This result implies that cis-trans distortion significantly reduces the square-net-derived states. Consequently, the states from the buffer layers, i.e., As(2,3) and La, should play a more important role in the physical properties of \ce{LaAgAs2}. 

Figure \ref{fig6}(b) shows the calculated band structure for \ce{LaAgAs2}. It can be seen that the bands cross $E_\mathrm{F}$ at $\Gamma$ point, X point, and the S$-$Y direction, leading to a Fermi surface as shown in Fig. \ref{fig6}(f). The Fermi surface consists of two 2D hole pockets at the $\Gamma$ point, a quasi-1D electronic pocket at the X point, and two tiny 3D electronic pockets along the S$-$Y direction. Further details for each pocket are displayed in Supplementary Fig. S7 \cite{SM}.  

The calculated electronic structure of \ce{LaAgAs2} is in good agreement with that detected by the ARPES measurements except for the size of the 1D electron pocket at X point, an additional tiny pocket inside the 1D electron pocket, and the absence of 3D electron pockets along the S$-$Y direction. As the electron pockets are derived from the As1-orbitals within the cis-trans chains, they are very sensitive to the structure of these chains and that could be the origin of all discrepancies between the calculated and measured Fermi surfaces. 

To uncover the impact of cis-trans distortion on the electronic structure, we also calculate the electronic structure of the hypothetical tetragonal \ce{LaAgAs2}, with cis-trans chains artificially arranged to a checkerboard-like square net [Fig. \ref{fig6}(g)]. Because the unit cell of \ce{LaAgAs2} is related to the tetragonal \ce{HfCuSi2}-type structure by $\sqrt{2}a_\mathrm{T} \times \sqrt{2}a_\mathrm{T} \times 2c_\mathrm{T}$, the electronic structure of \ce{LaAgAs2} is folded as compared to that of $AMPn_2$, where the relationship between the folded and unfolded Brillouin zone is illustrated in Fig. \ref{fig6}(e). Both the folded [Figs. \ref{fig6}(c) and \ref{fig6}(g)] and unfolded [Figs. \ref{fig6}(d) and \ref{fig6}(h)] versions for the electronic structure of the hypothetical tetragonal \ce{LaAgAs2} are shown to facilitate the comparison with the orthorhombic \ce{LaAgAs2} and typical $AMPn_2$, respectively.  From Figs. \ref{fig6}(b) and \ref{fig6}(c), it can be seen that the orthorhombic structure distortion mainly modifies the bands around the zone boundary with the As1 character, with the hole bands around the zone center remaining intact, in good agreement with the ARPES measurements. Specifically, cis-trans distortion changes the 2D linear Dirac bands into quasi-1D parabolic bands, consistent with the theoretical prediction by Tremel \emph{et. al.} \cite{tremel_square_1987}.

The transport properties of \ce{LaAgAs2} are similar to those in $AMPn_2$ compounds, which are dominated by the $Pn$ square net-derived Dirac bands. However, from the above analysis, the square net-derived bands initially expected to account for the quantum transport phenomena are transformed into quasi-1D trivial bands by the cis-trans distortion. Therefore, bands around the $\Gamma$ point with primarily As(2,3) orbital character should be responsible for the observed quantum transport phenomena. Indeed, the $F_\beta$ with a higher frequency can be assigned to the inner hole pockets around the $\Gamma$ point. However, no Fermi pocket responsible for the lower frequency $F_\alpha$ can be identified at present.

From Figs. \ref{fig6}(b-d), it can be seen that the linear conduction bands nearly touch the valence bands at the $\Gamma$ point as observed in Heusler topological insulators \cite{chadov_tunable_2010,lin_half-heusler_2010}, indicating the proximity of a topological quantum transition. Upon a close inspection of the band structure without considering SOC at the $\Gamma$ point [Fig. \ref{fig7}(a)], the valence and conduction bands are separated by a tiny gap about $\sim$0.6 meV, which is much smaller than the strength of the SOC. Since the system possesses inversion symmetry, we can calculate the parity values of the occupied states at the eight time-reversal invariant momenta (TRIM) points according to the Fu-Kane formula ~\cite{prb7604} (see Supplementary Table S1 for details \cite{SM}). The results indicate that the system exhibits $Z_{2}$ topologically non-trivial characteristics when it is fully occupied by valence electrons (256 valence electrons). Therefore, it can be concluded that the SOC induces a band inversion between the valence and conduction bands and leads to a topological surface state (TSS) with spin-momentum locked Dirac-cone located in the gap between the valence and conduction bands [Fig. \ref{fig7}(b)] \cite{Kane2005,Noh2008a,Hsieh2008,Hasan2010,Pan2011c}. We further notice that two valence bands cross around $E_\mathrm{F}$, resulting in a bulk 3D Dirac cone [Fig. \ref{fig7}(b)].

Therefore, both the TSS and topological Dirac semimetal (TDS) states coexist in \ce{LaAgAs2}, as illustrated in Fig. \ref{fig7}(b). The multiple topological states observed in \ce{LaAgAs2} are similar to those in \ce{LiFe_{1-x}Co_xAs}, \ce{FeTe_{1-x}Se_x} \cite{zhang_multiple_2019}, where both the TSS and TDS are identified around the $\Gamma$ point \cite{zhang_multiple_2019}. These results were also cross-validated through the application of topological quantum chemistry methods, validating the predictive capabilities of topological quantum chemistry theory in the discovery of topological materials \cite{bradlyn_topological_2017,vergniory_complete_2019,vergniory_all_2022}. 

Because both the TSS and TDS states are located well above the Fermi level, they cannot be captured by ARPES measurements (see the ARPES section). Furthermore, due to the highly delocalized nature of the La's $d$ orbitals and As1's $p$ orbitals, generating well-localized Wannier orbitals for subsequent surface state calculations via the surface Green's function method \cite{mpls,mpls2} proved challenging. The quantum oscillation measurements indicate that the $F_\alpha$ bands show the characteristics expected by a nontrivial topological state, i.e., the quasi-2D Fermi surface with small size and small effective mass. Since the TDS cone is located closer to $E_\mathrm{F}$ than the TSS cone, $F_\alpha$ likely originates from the TDS cone.

\section{Discussion}

Our calculations have clearly demonstrated that \ce{LaAgAs2} is a topological material with multiple topological states, similar to iron-based superconductors, but differs significantly from $AMPn_2$ sister compounds. Our magnetotransport experiments also suggest that the $F_\alpha$ oscillation originates from nearly massless carriers, most likely from the bulk TDS cone. These results can be well understood from the perspective of chemical bonds. 

From Fig. \ref{fig7}(a-b) and orbital-resolved DOS shown in Supplementary Figs. S5 and S6 \cite{SM}, it can be seen that it is As(2,3)-$p_x$/$p_y$, La-$d_{xy}$, and Ag-$d_{xz}/d_{yz}$ that dominate the dispersive bands around $\Gamma$, indicative of primarily in-plane bonding interactions. This result differs from the interlayer bonding in iron-based superconductors \cite{zhang_multiple_2019, shi_FeTeSe_2017}, suggesting a distinct origin of the multiple topological states. 

As illustrated in Fig.\@ \ref{fig7}(c), the As(2,3) atoms, with four As(2,3) neighbors in the same layer,  are sandwiched by four La(1,2) and four Ag(1,2) atoms. Notice that we do not distinguish the detailed atomic position of La(1,2), Ag(1,2), and As(2,3) atoms due to the pseudo-tetragonal nature of the La and [AgAs] layers. The distance between As(2,3) atoms is $\sim$4.12 {\AA}, too long to form the square-net-derived bands as predicted by Hoffmann \emph{et. al.} \cite{hoffmann_making_1985}. The interatomic distance of As(2,3)-La is $\sim$0.95 {\AA}, much shorter than $\sim1.85$ {\AA} between As(2,3) and Ag layers, indicating that La atoms play a more important role in the in-plane bonding interactions, which is corroborated by the higher orbital weight of La in the valence and conduction bands [Supplementary Figs. S5 and S6 \cite{SM}]. This result suggests that the puckered [LaAs(2,3)] layer with an in-plane checkerboard configuration can be treated as an entirety, i.e., a building block [Fig. \ref{fig7}(d)]. Therefore, we replot the crystal structure of \ce{LaAgAs2} in Fig. \ref{fig7}(e), emphasizing the important role played by the [LaAs] layer in the topological electronic structure. Due to their small electronegativity, rare earth elements are typically regarded as spacer atoms that donate electrons to the system, as reported in iron-based superconductors \cite{liu_role_2023,chen_iron-based_2014}. Our results indicate that the structural motif can be rearranged in specific materials and identify a new structural motif, [LaAs], capable of hosting multiple topological states. Since the puckered $[REPn]$ layer is a common structural motif among the square-net-based structures, such as \ce{HfCuSi2}, \ce{ThCr2Si2}, PbFCl, and ZrCuSiAs, all of which exhibit great structural diversity \cite{klemenz_topological_2019,just_coordination_1996,pottgen_materials_2008}. Thus, our results can serve as a guideline for discovering new topological materials.

Finally, we want to point out to an interesting possibility that the extremely small orthorhombicity in LaAgAs$_2$ and its sister compounds, of the order of $\sim5\times10^{-4}$, could make these materials potentially very sensitive to the uniaxial strain \cite{Steppke2017,Kundu2024}. A routinely achievable strain of $\sim1\%$, $1-2$ orders of magnitude larger than the orthorhombicity, might not only re-establish tetragonal order, but it might be able to drive the system to an uncharted territory with easily tunable electronic structure, where the topological character could be turned on and off on demand.

\section{Methods}

\textbf{Single crystal growth}. Single crystals of \ce{LaAgAs2} were grown by the self-flux method using excess Ag and As as flux \cite{mondal_magnetocrystalline_2018}. The starting materials of La chunks, Ag grains, and As lumps were weighted according to the ratio of \ce{LaAg_{36}As_{17}}, which were mixed and loaded into an alumina crucible. The crucible was sealed in an evacuated quartz tube and then slowly heated to \SI{970}{\degreeCelsius} in a box furnace. After dwelling at \SI{970}{\degreeCelsius} for 10 h, the sample was cooled to \SI{750}{\degreeCelsius} at a rate of \SI{2}{\degreeCelsius}/h to grow single crystals. Shiny centimeter-sized single crystals with a typical size of $10 \times 5 \times 0.5$ \unit{mm^3} can be obtained by decanting the excess flux using a centrifuge. 

\textbf{Structure characterizations}. Single-crystal XRD measurements were performed on a high-flux, high-resolution, rotating anode Rigaku Synergy-DW (Mo/Ag) diffractometer using Mo $K_\mathrm{\alpha}$ radiation ($\lambda$ = 0.7107 {\AA}). The system is equipped with a background-less Hypix-Arc150$^{\circ}$ detector, which guarantees minimal reflection profile distortion and ensures uniform detection conditions for all reflections. All samples were measured to a resolution better than 0.5 {\AA} and with the beam divergence set to 5 mrad.\@ The samples exhibited no mosaic spread and no additional reflections from secondary phases, highlighting their high quality and allowing for excellent evaluation using the latest version of the CrysAlisPro software package \cite{CrysAlis}.\@ The crystal structure of the system was refined using JANA2006 \cite{Vaclav_229_2014}.\@ Out-of-plane x-ray diffraction patterns were collected using a PANalytical powder diffractometer (Cu $K_{\alpha}$ radiation). Laue diffraction patterns were taken in a backscattering geometry with white incident light exposed from the out-of-plane direction.

\textbf{Magnetization and transport measurements}. Both the magnetization and magnetotransport measurements were performed in a Quantum Design DynaCool Physical Properties Measurement System (PPMS-14T). The VSM option was used for magnetization measurements, and the DC delta mode with a Keithley 6221 current source and a Keithley 2182A nanovoltmeter was used to record the transport data. In-plane electrical resistivity $\rho_{xx}$ and Hall resistivity $\rho_{xy}$ were measured simultaneously on a polished single crystal with the six-probe configuration. To eliminate the mixture of resistivity and Hall signals arising from the geometrical factor of the electrical contacts, the $\rho_{xx}$ and $\rho_{xy}$ data shown in this paper were corrected according to the following two formulas: $\rho_{xx}(B)=\left[\rho_{xx}(+B)+\rho_{xx}(-B)\right]/2$ and  $\rho_{xy}(B)=\left[\rho_{xy}(+B)-\rho_{xy}(-B)\right]/2$. The out-of-plane resistivity $\rho_{zz}$ was measured using a standard four-probe method, where the sample was prepared by cutting a thick single crystal into a needle-shaped bar with the long axis along the crystalline $b$ axis (out-of-plane direction).  

\textbf{ARPES measurements}. The ARPES experiments were performed at the Electron Spectro-Microscopy (ESM) 21-ID-1 beamline of the National Synchrotron Light Source II and at BL-10.0.1.2 of the Advanced Light Source. Both beamlines are equipped with a Scienta DA30 electron analyzer, with base pressure $\sim2\times10^{-11}$ mbar. The total energy resolution was $\sim 10$ and $\sim 20$ meV at ESM and BL-10.0.1.2, respectively. The photon spot size on the sample is estimated to be $\sim2-3$ $\mu$m and $\sim50$ $\mu$m at ESM and BL-10.0.1.2, respectively. The angular resolution was $\sim$ \ang{0.1}, and $\sim$ \ang{0.3} along the slit and perpendicular to it, respectively, in both facilities. The polarization of the light was always linear and horizontal.

\textbf{First-principles calculations}. First-principles calculations were conducted using the Vienna $ab$-initio Simulation Package (VASP) within the framework of density functional theory (DFT) \cite{kresse_ab_1993}. The exchange-correlation interactions were described using the Perdew-Burke-Ernzerhof (PBE) functional under the generalized gradient approximation (GGA) \cite{perdew_generalized_1996}. Projector-augmented-wave (PAW) pseudopotentials were employed with a plane-wave basis set energy cutoff of 400 eV \cite{kresse_efficient_1996,kresse_efficiency_1996}. The calculations used an orthorhombic crystal structure based on our experimental lattice parameters of $a_\mathrm{o}$ = 5.8350(1) {\AA},\@ $b_\mathrm{o}$ = 21.2926(4) {\AA},\@ and $c_\mathrm{o}$ = 5.8306(1) {\AA}.\@ For structural optimization, all atomic positions were fully relaxed until the residual forces on each atom were less than 0.001 eV/{\AA}, with convergence achieved within an energy threshold of \num{e-5} eV. A $\Gamma$-centered Monkhorst–Pack $k$-point mesh of \numproduct{9 x 9 x 2} was utilized to sample the first Brillouin zone. The electronic band structure was computed with and without the inclusion of spin-orbital coupling (SOC) to capture the relativistic effects. The Fermi surface at different energy levels was analyzed and refined using the iFermi software \cite{ganose_ifermi_2021}. The hypothetical tetragonal \ce{LaAgAs2} was calculated with the same method as the orthorhombic one. The tetragonal structure is obtained by artificially setting $a = b = a_\mathrm{o}$ and $c = b_\mathrm{o}$, and arranging the cis-trans chains into the square net fashion. Parity analysis is conducted using the irvsp code \cite{GAO2021107760}.

\vspace{12pt}
\noindent\textbf{Data availability:} The datasets generated and/or analyzed during the current study are not publicly available due to legal restrictions preventing unrestricted public distribution, but are available from the corresponding authors upon reasonable request.

\noindent\textbf{Competing interests:} The authors declare no competing financial or non-financial interests.

\noindent\textbf{Acknowledgments:} We thank Meng Zhang, Dashuai Ma, and Tao Wu for their helpful discussions. We would like to thank Guiwen Wang and Yan Liu at the Analytical and Testing Center of Chongqing University for their assistance with transport measurements. We gratefully acknowledge Siegmar Roth and Andre Beck at the Institute for Quantum Materials and Technologies, Karlsruhe Institute of Technology, for their technical support with x-ray diffraction.

\noindent\textbf{Funding:} The work at Chongqing University was supported by the National Key Research and Development Program of China (Grant No. 2025YFA1411300), the Natural Science Foundation of Chongqing, China CSTC (Grant No. CSTB2024NSCQ-QCXMX0002), the Fundamental Research Funds for the Central Universities, China (Project No. 2025CDJ-IAISYB-031), and the National Natural Science Foundation of China (Grant Nos. 12474142, 12104072). Y.C. is supported by the Open Research Fund of the Pulsed High Magnetic Field Facility (Grant No. WHMFC2024007), Huazhong University of Science and Technology. This research used resources of the Advanced Light Source, a U.S. DOE Office of Science User Facility under contract no. DE-AC02-05CH11231 and of the National Synchrotron Light Source II, a U.S. DOE Office of Science User Facility under Contract No. DE-SC0012704.
T.V. acknowledges the support from the Red guipuzcoana de Ciencia, Tecnología e Innovación – Gipuzkoa NEXT 2023 from the Gipuzkoa Provincial Council under Contract No. 2023-CIEN-000046-01 and the Center for Advanced Laser Techniques (CALT), grant No. KK.01.1.1.05.0001.

\noindent\textbf{Author Contributions:} T.V. and A.W. conceived the project. Y.L. and A.W. grew the single crystals. T.L., X.Y., Z.S., and H.F. performed the DFT calculations. N.M. and M.M. performed the single-crystal XRD measurements. A.V.F., A.K.K., T.Y., E.V., and T.V. performed ARPES measurements. Y.L., X.W., L.Z., M.H., Y.C., and X.Z. performed the transport measurements. M.M., T.V., and A.W. wrote the manuscript with inputs from all co-authors. All authors read and approved the final manuscript.

\section{References}

\newpage

\begin{table}
\caption{Crystallographic results of LaAgAs$_2$ as determined from single-crystal x-ray diffraction at 300, 200, and 80 K.\@ For all temperatures, the structure was refined in the orthorhombic space group $Pbcm$.\@ The lattice parameters, $a$, $b$, $c$, and volume $V$ are shown together with the Wyckoff positions of the atoms, and the equivalent atomic displacement parameters $U_{\rm eq}$.\@ The ADPs were refined anisotropically, but due to space limitations, only the $U_{\rm eq}$ are listed in the Table.\@ $TW$ denotes the twinning ratio of the sample. The errors shown are statistical errors from the refinement.
}

\label{tab_cryst}

\renewcommand{\arraystretch}{0.66}
\small
\scalebox{0.8}{
\begin{ruledtabular}

    \begin{tabular} {c c c c c }

       & Temperature & 300 K & 200 K & 80 K \\ \hline
       & SG &  $Pbcm$ &  $Pbcm$  & $Pbcm$ \\   
        & $a$ (\AA)&  5.8350(1) & 5.8266(1) &  5.8198(1) \\  
        & $b$ (\AA)&  21.2926(4)&  21.2585(5)&  21.2284(5) \\  
          & $c$ (\AA)& 5.8306(1) &  5.8206(1)& 5.8172(1) \\ 
        & $V$ (\AA$^3$)& 724.4 & 721.0  &  718.7 \\  \hline 
La1 &  Wyck.&   4d  &  4d    & 4d    \\   
&  $x$  & 0.01295(6) &  0.01300(6)& 0.01297(6) \\
&  $y$   &  0.38575(2)&  0.38571(2)& 0.38567(2) \\
&  $z$   & 1/4 &  1/4 & 1/4 \\
&  $U_{\rm eq}$&  0.00848(11)&   0.00617(11)& 0.00398(10) \\
La2 &  Wyck.&  4d  &  4d  & 4d   \\   
&  $x$  &  0.51428(6)  &  0.51439(6)& 0.51434(6)  \\ 
&  $y$  &  0.11949(2) &  0.11952(1)  & 0.11956(1) \\
&  $z$   &  1/4 &  1/4  & 1/4   \\
&  $U_{\rm eq}$&  0.00780(11)&  0.00567(11)&  0.00388(10) \\
Ag1 &  Wyck.&  4c &  4c  & 4c   \\   
&  $x$  & 0.26271(8) &  0.26286(8)&   0.26292(7)   \\ 
&  $y$   &  1/4  &  1/4  & 1/4  \\
&  $z$    &  0 &  0   & 0  \\
&  $U_{\rm eq}$&  0.01164(21) & 0.00786(20)   & 0.00473(18) \\
Ag2    &  Wyck.&  4c  &  4c   & 4c \\   
&  $x$   & 0.76594(8) &  0.76612(8)& 0.76601(8)     \\ 
&  $y$    &  1/4  &  1/4  & 1/4  \\
&  $z$     & 0    &  0   & 0  \\
& $U_{\rm eq}$&  0.01226(20) &  0.00889(20)& 0.00501(18) \\
As1 &  Wyck.&   8e  &  8e   & 8e  \\   
&  $x$    & 0.28383(7)  &  0.28405(8)& 0.28392(7)  \\ 
&  $y$   & 0.00118(3) &  0.00116(3)  &  0.00119(3)  \\
&  $z$     &  0.03005(7)  &  0.03016(7)& 0.03015(3)  \\
& $U_{\rm eq}$&  0.01246(10) &  0.00941(9)&  0.00665(9) \\
As2 &  Wyck.& 4d  &  4d   & 4d  \\   
 &  $x$   & 0.51406(9)   &  0.51411(10)& 0.51411(10) \\ 
&  $y$   &   0.33591(3)  &  0.33584(3)& 0.33580(3)  \\
&  $z$     &  1/4  &  1/4 &  1/4  \\
& $U_{\rm eq}$&  0.00892(18) &  0.00636(19)& 0.00417(18) \\
As3 &  Wyck.& 4d  &  4d   &  4d \\   
&  $x$    & 0.01425(10)  &  0.01439(10)& 0.01440(10) \\ 
&  $y$   & 0.16336(3)  &  0.16346(3)&  0.16354(3)  \\
&  $z$     &  1/4  &  1/4 &  1/4  \\ 
& $U_{eq}$&  0.00906(18) & 0.00668(19)& 0.00441(18) \\ \hline
& $TW$ (\%) &  58/42  &  58/42 &  58/42  \\
& wR$_2$ (\%) &  6.64  &  6.66 &  6.62  \\
 & R$_1$ (\%) &  3.09  &  3.13  & 3.19    \\
& GOF &  1.85  &  1.81  &  1.85 \\ 
    \end{tabular}
\end{ruledtabular}
}
\end{table}

\begin{figure}
\includegraphics[width=8.5cm]{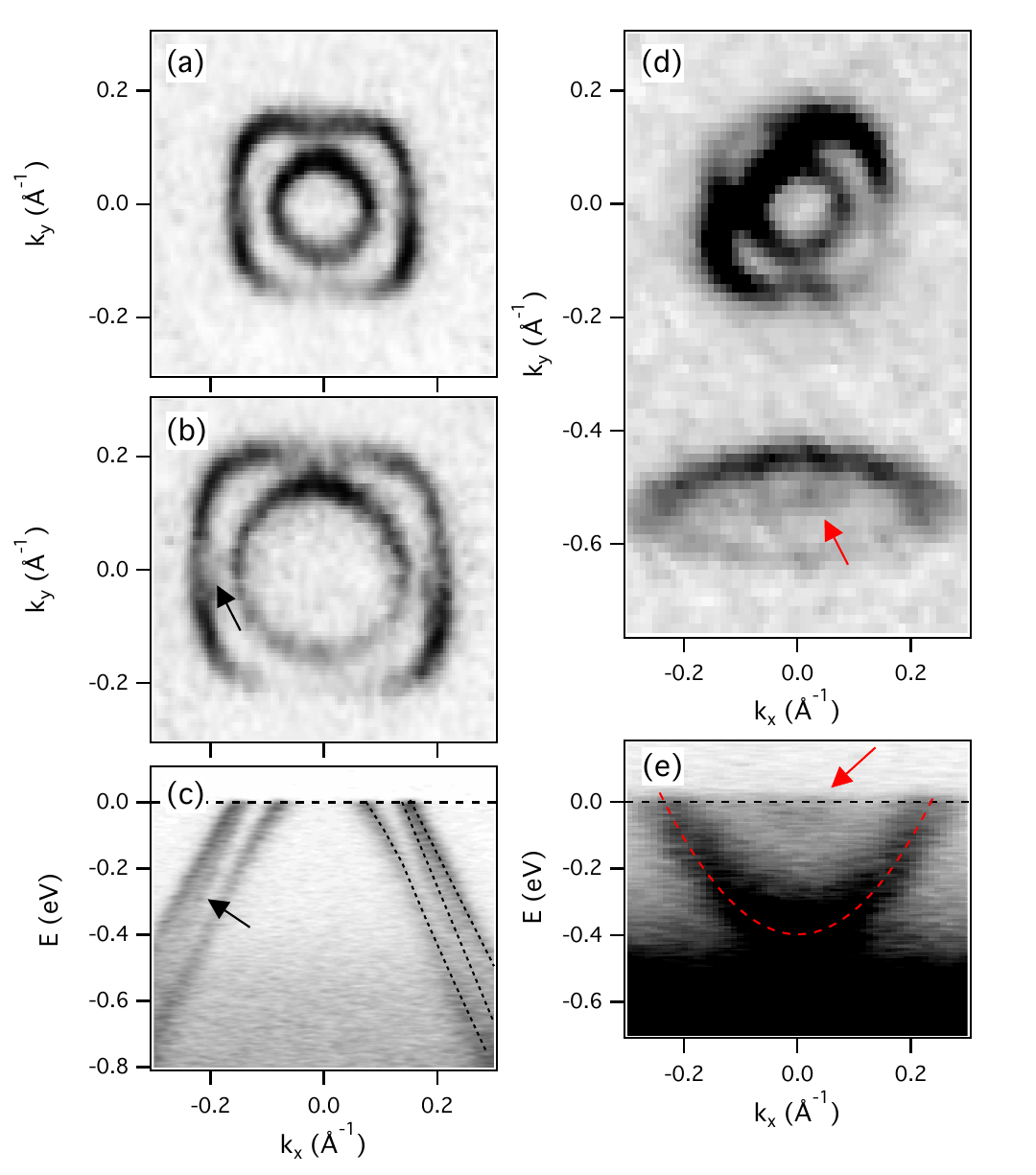}
\caption{High-resolution electronic structure of the Fermi pockets. (a) The Fermi surface and (b) the intensity at $E=-0.2$ eV taken at $h\nu=100$ eV (c) Band structure along the $k_y=0$ line from (a). The dotted lines at $k_x>0$ indicate the three resolved hole bands. The black arrows point to the splitting of the outer hole doublet. (d) The Fermi surface taken at $h\nu=70$ eV (e) Band structure along the $k_y=\pi/c$ line from (d). The red arrows indicate the small electron pocket inside the main one. The red dashed curve represents the parabolic fit to the MDC-derived dispersion.}
\label{fig5}
\end{figure}

\begin{figure*}
\includegraphics[width= \linewidth]{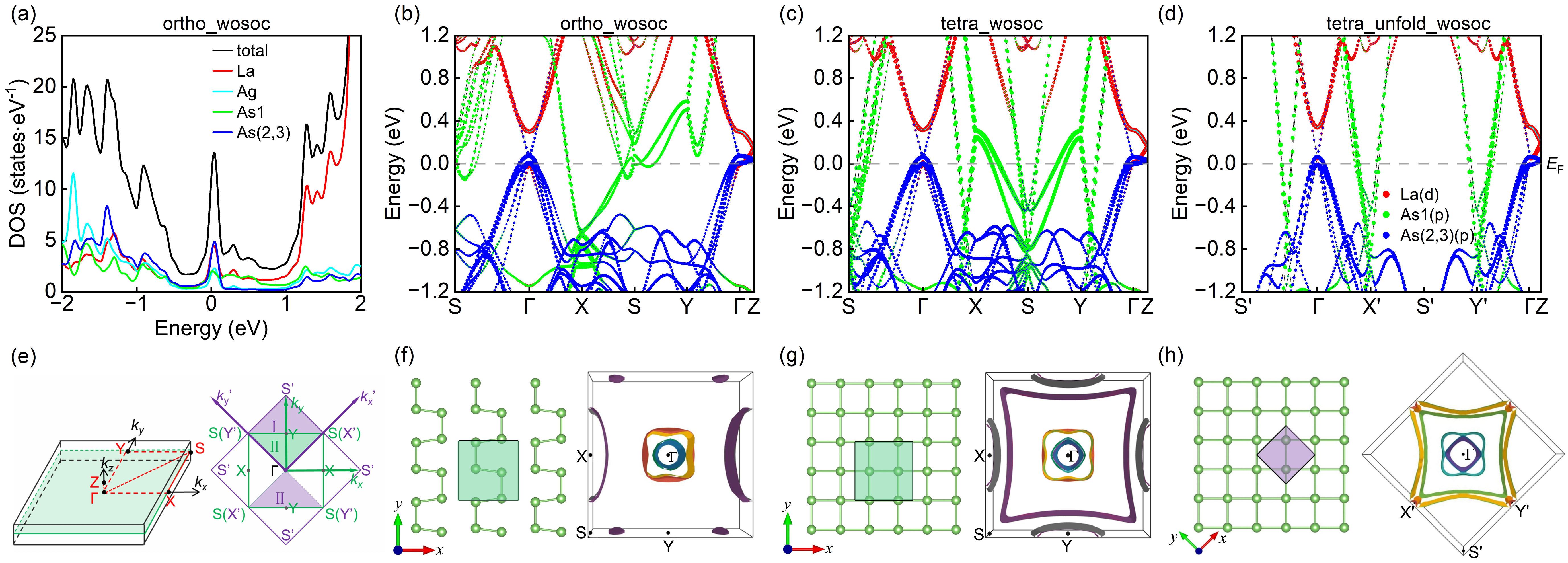}
\caption{The calculated electronic structure of \ce{LaAgAs2}. (a) Atom-specific partial density of states (DOS). The orbital-projected band structure without spin-orbit coupling for \ce{LaAgAs2} with pristine orthorhombic structure ($a_\mathrm{o} \times b_\mathrm{o} \times c_\mathrm{o}$) (b), hypothetical tetragonal structure ($a_\mathrm{o} \times a_\mathrm{o} \times b_\mathrm{o}$ or $\sqrt{2}a_\mathrm{T} \times \sqrt{2}a_\mathrm{T} \times 2c_\mathrm{T}$) (c), and unfolded hypothetical tetragonal structure ($a_\mathrm{T} \times a_\mathrm{T} \times 2c_\mathrm{T}$) (d), where $a_\mathrm{o} = \sqrt{2}a_\mathrm{T}$ and $b_\mathrm{o} = 2c_\mathrm{T}$, $a_\mathrm{T}$ and $c_\mathrm{T}$ refer to a \ce{HfCuSi2}-type unit cell. The correspondence of the high symmetry points between the folded (green) and unfolded (purple) BZ is illustrated in (e). (f-h) display the top view of the crystal structure of the As1 layer (left) and the Fermi surfaces (right), where the green/purple shaded square on the left panel indicates the unit cell corresponding to the folded/unfolded BZ. }
\label{fig6}
\end{figure*}

 \begin{figure}
\includegraphics[width=4in]{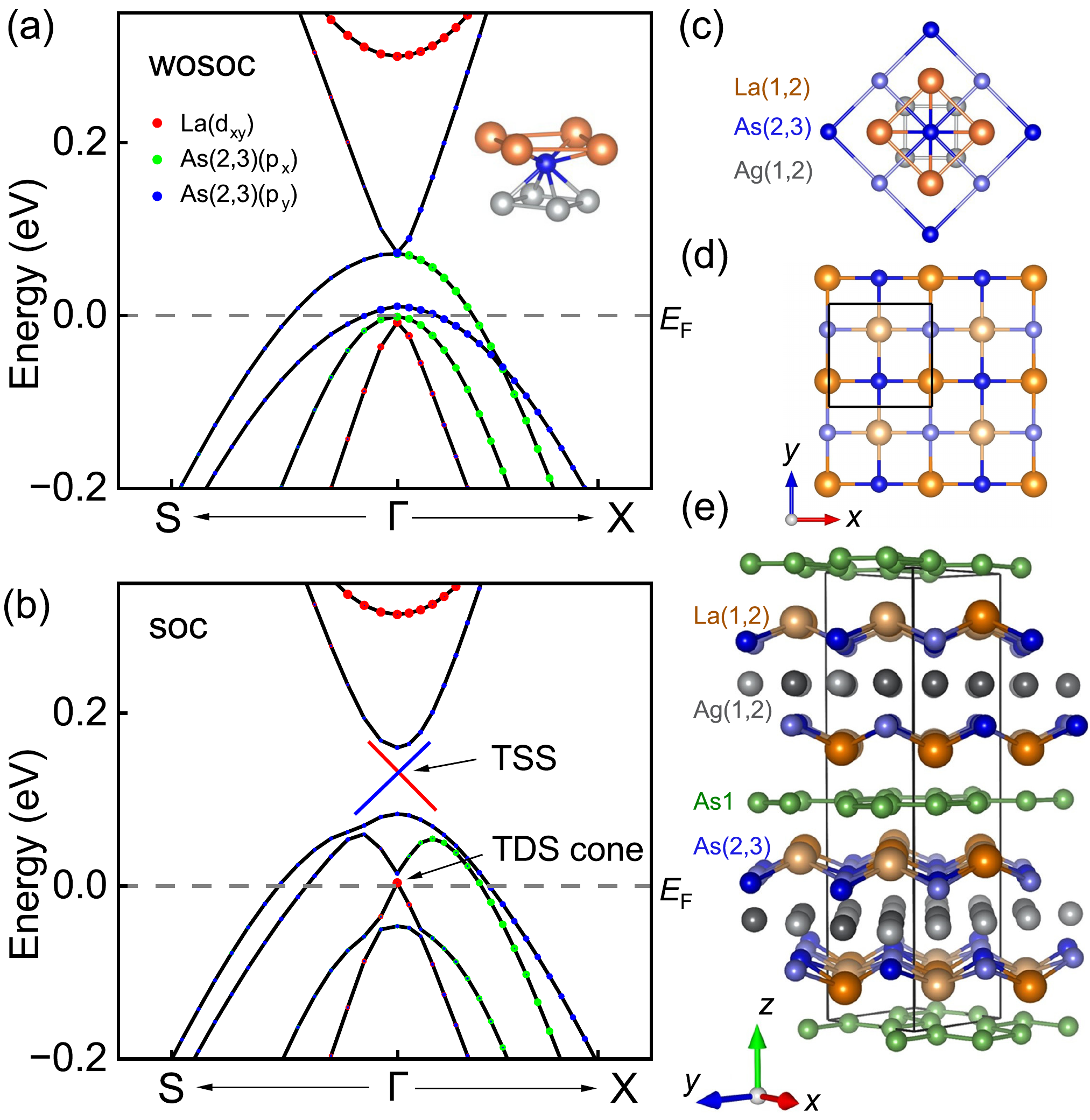}
\caption{The enlarged view of the calculated band structure around the $\Gamma$ point. Panels (a) and (b) show the bands without and with considering spin-orbit coupling, respectively. The sketches of the spin-polarized topological surface state (TSS) and topological Dirac semimetal state (TDS) are marked in (b). (c) An As(2,3) atom with its nearest neighbors. (d) Top view of the puckered [LaAs(2,3)] layer, where the atoms are arranged in the checkboard square net fashion. The crystal structure of \ce{LaAgAs2} is rearranged from Fig. \ref{fig1}(a) into panel (e), emphasizing the role of the [LaAs(2,3)] layer played in the topological electronic structure.}
\label{fig7}
\end{figure}

\end{document}